\documentclass[final,3p,times,twocolumn]{elsarticle}
\usepackage{amssymb, amsmath}
\usepackage{lineno,hyperref}
\usepackage{siunitx}
\usepackage{comment}
\usepackage{multirow}
\usepackage{here}
\modulolinenumbers[5]
\journal{Journal of \LaTeX\ Templates}

\begin{document}

\begin{frontmatter}

\title{Development of a Low-noise Front-end ASIC for CdTe Detectors}
\author[Todai,IPMU]{Tenyo Kawamura\corref{mycorrespondingauthor}}
\cortext[mycorrespondingauthor]{Corresponding author}
\ead{tenyo.kawamura@ipmu.jp}
\author[IPMU]{Tadashi Orita}
\author[IPMU]{Shin'ichiro Takeda}
\author[ISAS,IPMU]{Shin Watanabe}
\author[ISAS]{\\Hirokazu Ikeda}
\author[IPMU,Todai]{Tadayuki Takahashi}

\address[Todai]{Department of Physics, University of Tokyo, 7-3-1 Hongo Bunkyo, Tokyo 113-0033, Japan}
\address[IPMU]{Kavli Institute for the Physics and Mathematics of the Universe (WPI), University of Tokyo, 5-1-5 Kashiwanoha, Kashiwa, Chiba 277-8583, Japan}
\address[ISAS]{Institute of Space and Astronautical Science, Japan Aerospace Exploration Agency (ISAS/JAXA), 3-1-1 Yoshinodai, Chuo-ku, Sagamihara, Kanagawa 252-5210, Japan}

\begin{abstract}
We present our latest ASIC, which is used for the readout of 
Cadmium Telluride double-sided strip detectors (CdTe DSDs) and 
high spectroscopic imaging.
It is implemented in a $0.35\,\si{\micro m}$ CMOS technology (X-Fab XH035), 
consists of 64 readout channels and 
has a function that performs simultaneous AD conversion for each channel.
The equivalent noise charge of $54.9\,\mathrm{e}^{-} \pm 11.3\,\mathrm{e}^{-}$ (rms)
is measured without connecting the ASIC to any detectors.
From the spectroscopy measurements using a CdTe single-sided strip detector,
the energy resolution of $1.12\,\si{keV}$ (FWHM) is obtained at $13.9\,\si{keV}$, and 
photons within the energy from $6.4\,\si{keV}$ to $122.1\,\si{keV}$ are detected.
Based on the experimental results, we propose a new low-noise readout architecture 
making use of a slew-rate limited mode at the shaper followed by a peak detector circuit.
\end{abstract}
\begin{keyword}
ASIC; Low-noise; X-ray; Gamma-ray; CdTe; Analog front-end
\end{keyword}
\end{frontmatter}

\section{Introduction}
Imaging spectroscopy of photons from $10\,\si{keV}$ to a few hundreds $\si{keV}$ 
has a variety of applications 
in astronomy, medicine and industry.
The wide use of this imaging technique 
has driven the development of imagers based on CdTe detectors, 
since they have high absorption efficiency comparable with that of NaI and CsI, and 
the predominance of photoelectric absorption up to $\sim 250\,\si{keV}$ 
\cite{Takahashi_2001, Jones_2009, Meuris_2009}.

In the field of in-vivo molecular imaging, 
large detection area of $\sim 10\,\si{cm^2}$ is required as well as 
the energy resolution of $\sim 1\,\si{keV}$ and 
spatial resolution of $\sim 100\,\si{\micro m}$ 
in order to image multiple radioisotopes\cite{Takeda_2018}.
A double-sided strip detector is a promising solution 
in terms of its small number of readout channels compared to a pixel detector. 
However large capacitance at the input of the signal processing circuit 
degrades the noise performance.
It is therefore crucial to design a low-noise readout chip 
which operates under a large detector capacitance.

We present our latest ASIC named KW04H64 
which was designed for the readout of CdTe DSD 
having a detection area of $\sim 32\,\si{mm} \times 32\,\si{mm}$, 
a strip pitch of $250\,\si{\micro m}$, and 
a capacitance of $\sim 10\,\si{pF}$ for each channel.
The ASIC has been modified from our previous versions
(e.g. \cite{Ikeda_2006, Kishishita_2010}).

Section \ref{sec:asic_description} describes the signal processing architecture of the ASIC and 
its predicted performance based on simulation results. 
Section \ref{sec:experimental_results} reports the first ASIC measurements. 
In Section \ref{sec:noise_performance} and \ref{sec:dynamic_range_performance} 
we discuss the measured noise performance and dynamic range of the ASIC 
when not connected to a detector. 
In Section \ref{sec:spectroscopic_performance}, 
we evaluate the spectroscopic performance 
when the ASIC is connected to a CdTe and Silicon (Si) semiconductor detector, 
and in Section \ref{sec:new_low_noise_readout_architecture} 
we propose a new low-noise readout architecture.

\section{ASIC Description}
\label{sec:asic_description}
\begin{figure*}
	\begin{center}
		\includegraphics[width=\linewidth]
		{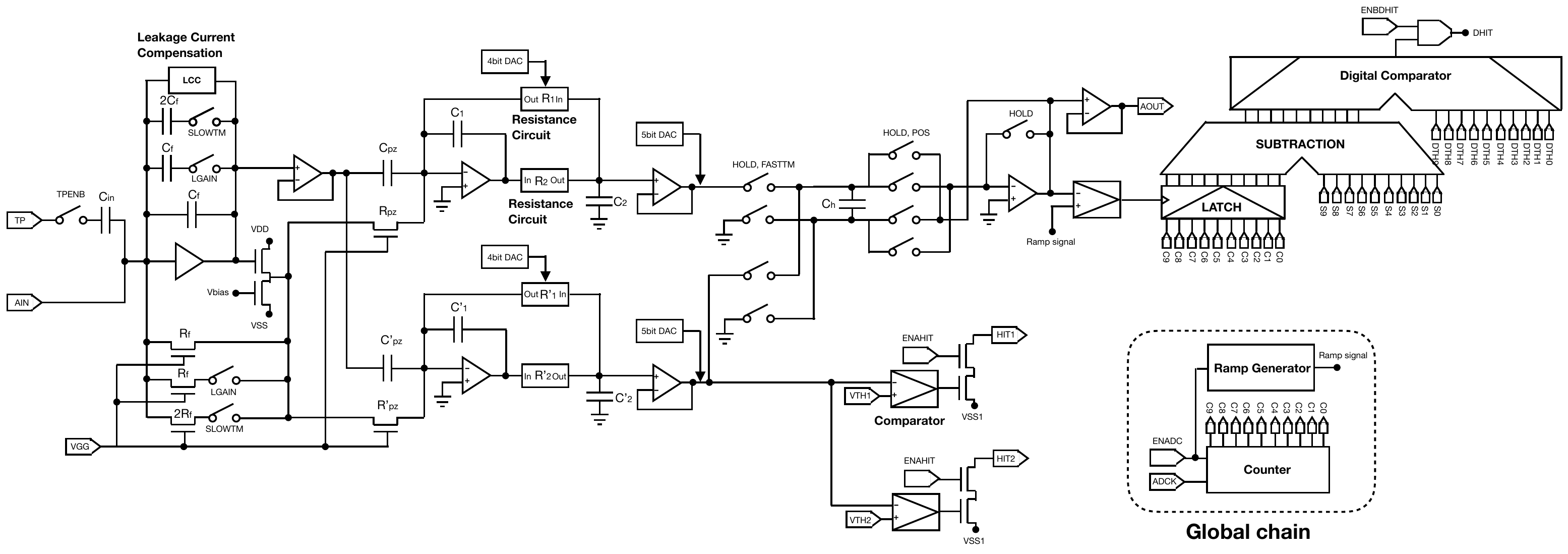}
		\caption{Schematic of the signal processing circuit 
		implemented in each readout channel.
		Typical values of resistors and capacitors are 
		$R_{\mathrm{f}}=5\,\si{G\Omega}$,
		$C_{\mathrm{f}}=0.032\,\si{pF}$, 
		$R_{\mathrm{pz}}=50\,\si{M\Omega}$,
		$C_{\mathrm{pz}}=3.2\,\si{pF}$, 
		$R_{\mathrm{1}}=R_{\mathrm{2}}=8\,\si{M\Omega}$,
		$C_{\mathrm{1}}=0.4\,\si{pF}$, 
		$C_{\mathrm{2}}=0.1\,\si{pF}$, 
		$R' _{\mathrm{1}}=R' _{\mathrm{2}}=0.8\,\si{M\Omega}$,
		$C' _{\mathrm{1}}=0.4\,\si{pF}$, 
		$C' _{\mathrm{2}}=0.1\,\si{pF}$ and 
		$C' _{\mathrm{h}}=0.8\,\si{pF}$.
		Note that the ramp signal generator and the counter used in the Wilkinson ADC 
		are implemented in another block in the ASIC whose outputs are provided to each channel.} 
		\label{fig:kw04h64_chain1a}
	\end{center}
\end{figure*}
\subsection{Overview}
\label{sec:overview}
The KW04H64 ASIC is implemented in a CMOS $0.35\,\si{\micro m}$ technology (X-Fab XH035).
The chip has 64 readout channels with self-trigger capability and 
it measures $7.12\,\si{mm} \times 8.03\,\si{mm}$.
The dual power supply line of $\pm 1.65\,\si{V}$ has been adopted 
to process the signal in either polarity, which is required for DSDs.
The main characteristics of the ASIC are summarized 
in Table \ref{tab:kw04h64_characteristics}.
\begin{table}
	\begin{center}
		\caption{Main characteristics of the ASIC KW04H64.} \vspace{3mm}
		{\footnotesize
			\begin{tabular}{ll}
			\hline
			Parameter 					 & Value \\ \hline
			Fabrication process 		 & X-Fab XH035 \\
			Chip size 					 & $7.12\,\si{mm} \times 8.03\,\si{mm}$ \\
			Number of channels  		 & 64 \\
			Power rail					 & $\pm 1.65\,\si{V}$ \\
			Polarity					 & Both \\
			Gain						 & $170\,\si{mV/fC}$\\
			Dynamic range				 & $\sim 6.0\,\si{fC}$\\
			\multirow{2}{*}{ENC}		 & $58.4\,\si{e^-} + 12.7\,\si{e^- /pF}$ (Fast shaper) \\
										 & $33.0\,\si{e^-} + 5.2\,\si{e^- /pF}$ (Slow shaper) \\
			\multirow{2}{*}{Peaking time}& $\sim 0.5\,\si{\micro s}$ (Fast shaper) \\
										 & $\sim 2.0\,\si{\micro s}$ (Slow shaper) \\
			Power consumption 			 & $2.1\,\si{mW/channel}$ \\ \hline
			\end{tabular}
		}
		\label{tab:kw04h64_characteristics}
	\end{center}
\end{table}

Figure \ref{fig:kw04h64_chain1a} shows the signal processing architecture of the ASIC. 
This architecture is the same as the previous KW04D64 chip version\cite{Harayama_2014}, 
with minor modifications concerning the circuit gain and includes additional functions.
It has a dual path topology after the charge sensitive amplifier (CSA).
As the names suggest, the ``timing branch'' is dedicated to measuring time, 
and the ``energy branch'' is dedicated to measuring energy.
The timing branch has
a pole-zero cancellation (PZC), 
a 2nd-order low-pass filter,
and two separate comparators which generate hit signals.
The energy branch has same architecture as that of the timing branch before comparators, 
followed by a sample-and-hold circuit 
and a Wilkinson ADC which enables the AD conversion for all the channels to happen simultaneously.
The PZC and the low-pass filter effectively work as the CR-RC shaper, 
which is called the ``fast shaper'' and ``slow shaper'' for the timing and energy branches 
and results from simulations give peaking times of 
$\sim 0.5\,\si{\micro s}$ and $\sim 2.0\,\si{\micro s}$, respectively.
The 5-bit and 4-bit DACs employed for the baseline adjustment
have a pair of output ports. 
The current supplied from one port of the DAC 
is designed to be recovered by the other port, 
which avoids any interference to adjacent circuits such as the buffer amplifier.
The sample-and-hold circuit is triggered by an external control block 
at a given time after a hit signal is generated.
The fast shaper signal from the timing branch 
can be selected as an input of the Wilkinson ADC in the energy branch.
This is beneficial for the performance measurement and the baseline adjustment of the fast shaper.

The signals are read out after the AD conversion at the Wilkinson ADCs for each event.
The ASIC has a sparse readout mode as well as a full readout mode.
In the sparse readout mode, 
only those channels carrying the signal data are read out in order to reduce the output data size.
Whether each channel carries the signal is judged by 
either the state of the hit signal (HIT1 in Figure\ref{fig:kw04h64_chain1a}) 
or the value of ADC subtracted by low frequency common mode noise\cite{Ikeda_1997}.

\subsection{Details of the analog signal processing}
\label{sec:details_of_analog_signal_processing}
The CSA employs a folded cascode scheme 
with a PMOS input transistor having $W/L=1440\,\si{\micro m}/1.2\,\si{\micro m}$ 
providing a transconductance of $\sim 3\,\si{mS}$ 
with a drain current of $\sim 160\,\si{\micro A}$.
The drain current can be changed 
by an externally located potentiometer. 
In this paper, the feedback capacitance at the CSA is fixed at $0.032\,\si{pF}$, 
although, depending on the photon energy, it can be set to 
$0.032\,\si{pF}$, $0.064\,\si{pF}$, $0.096\,\si{pF}$, and $0.128\,\si{pF}$.
The feedback resistor is implemented using an NMOS.
Its resistance is controlled by the gate voltage $\mathrm{VGG}$, 
which is generated by the internally located DAC.
Although it was switched off in this paper, 
a leakage current compensation circuit is implemented 
which can deal with up to $\sim\,1\,\si{nA}$ for either polarity 
(see \cite{Kishishita_2010, Krummenacher_1991} for details).

\begin{figure}
	\begin{center}
		\includegraphics[width=0.95\linewidth]
		{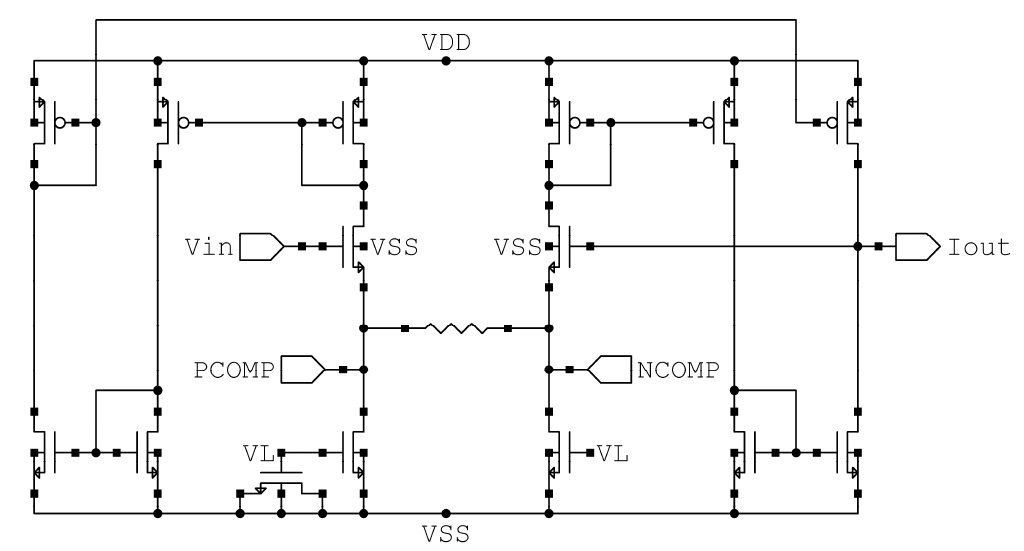}
		\caption{Schematic of the resistance circuit 
		implemented in the shaper. 
		Connected to a current DAC, 
		PCOMP and NCOMP can be used to calibrate a baseline coarsely.} 
		\label{fig:kw04h64_resistance_circuit}
	\end{center}
\end{figure}
\begin{figure}
	\begin{center}
		\includegraphics[width=0.75\linewidth]
		{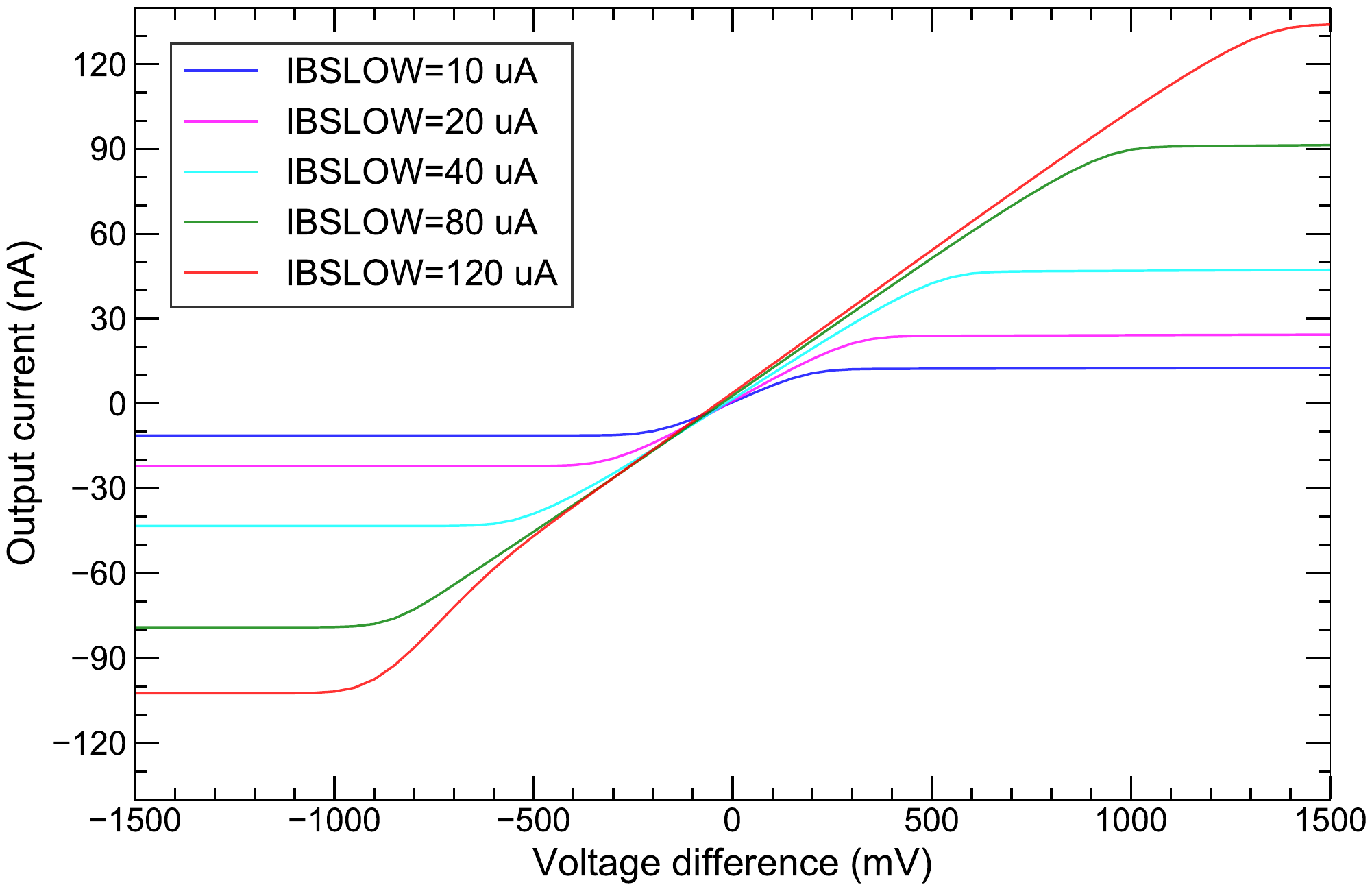}
		\caption{DC performance of the resistance circuit
		at the slow shaper predicted by simulations.} 
		\label{fig:kw04h64_VI8MLN_dc_sim}
	\end{center}
\end{figure}
At the shapers, the transconductors play an important role in the circuit performance. 
They are referred to as high resistance circuits 
having $\sim 8\,\si{M\Omega}$ for the slow shaper 
and $\sim 0.8\,\si{M \Omega}$ for the fast shaper.
The resistance circuits used in both the fast shaper and the slow shaper  
are implemented using the same architecture 
shown in Figure \ref{fig:kw04h64_resistance_circuit}, 
where two-stage current mirror scales down the current flowing into the passive resistor.
The bias current at the resistance circuits is controlled externally 
by changing the reference current 
in a similar manner to the drain current of the input PMOS at the CSA.
The reference current for the resistance circuits at the slow shaper 
is a key parameter in the paper, 
and is referred to hereafter as IBSLOW.
Note that the two resistance circuits at each shaper are biased in common 
to make the poles degenerated, i.e., to keep track of the critical damping condition.

The reference current determines the amount of output current. 
The maximum output current $I_{\mathrm{SR}}$ 
from the resistance circuit at the slow shaper is 
$I_{\mathrm{SR}} \sim (1/1000) \mathrm{IBSLOW}$.
Limiting the output current also limits the voltage difference 
under which the resistance circuit behaves as a linear resistor. 
The maximum voltage difference $V_{\mathrm{res, max}}$ is expressed as 
$V_{\mathrm{res, max}}=R_{\mathrm{res}}I_{\mathrm{SR}}$ 
where $R_{\mathrm{res}}$ is its resistance, 
i.e., $\sim 8\,\si{M\Omega}$ for the slow shaper.
When the voltage difference is larger than $V_{\mathrm{res, max}}$, 
the resistance circuit switches to the slew-rate limited mode, 
where it can only source or sink a maximum amount of current given by $I_{\mathrm{SR}}$.
Figure \ref{fig:kw04h64_VI8MLN_dc_sim} shows 
the DC performance of the resistance circuit at the slow shaper predicted by simulations
where the output voltage is fixed at $0\,\si{V}$, 
meaning the voltage difference described in the figure represents the input voltage.
It has been implied 
that the slew-rate limited mode 
contributes to better noise performance\cite{Sato_2009, Sato_2011}.

According to simulations, 
the maximum signal to be processed is expected to be $\sim 6\,\si{fC}$ 
with the gain of $\sim 170\,\si{mV/fC}$, 
while the noise performance is expected to be 
$\mathrm{ENC}=58.4\,\si{e^-} + 12.7\,\si{e^- /pF}$ and
$\mathrm{ENC}=33.0\,\si{e^-} + 5.2\,\si{e^- /pF}$ 
for the fast shaper and the slow shaper respectively.

	\begin{table}
		\begin{center}
			\caption{Main characteristics of the CdTe detector.} \vspace{3mm}
				{\footnotesize
				\begin{tabular}{ll}
				\hline
				Parameter   &Value \\ \hline
				Manufacturer&ACRORADO \\
				Type		&Schottky CdTe diode \\
				Size		&$4\,\si{mm} \times 4\,\si{mm}$ \\
				Thickness	&$1\,\si{mm}$ \\
				\multirow{2}{*}{Pt side}&16 strip electrodes \\
										&$250\,\si{\micro m}$ pitch \\  
				In side		&1 plain electrode \\ \hline
				\end{tabular}
				}
			\label{tab:cdte_detector}
		\end{center}
		\begin{center}
			\caption{Main characteristics of the Si detector.} \vspace{3mm}
				{\footnotesize
				\begin{tabular}{ll}
				\hline
				Parameter   &Value \\ \hline
				Manufacturer&Hamamatsu Photonics \\
				Type		&PN Si diode \\
				Size		&$12.8\,\si{mm} \times 12.8\,\si{mm}$ \\
				Thickness	&$300\,\si{\micro m}$ \\
				\multirow{2}{*}{P side}&32 strip electrodes \\
										 &$400\,\si{\micro m}$ pitch \\ \hline 
				\end{tabular}
				}
			\label{tab:si_detector}
		\end{center}
	\end{table}
\section{Experimental Results}
\label{sec:experimental_results}
	\subsection{Experimental Setup}
	The ASIC performance was evaluated 
	both without a detector and with Si and CdTe detectors. 
	In the setup without a detector, 
	an ASIC was placed in a QFP ceramic package on a test board and 
	test pulses generated internally were used as the input signals.
	The analog input pads of 56 readout channels are floated, 
	while for 8 channels each pad is connected to a pin on the test board.
	Due to the difference between these two types of configurations, 
	56 channels were used for the performance evaluation.
	In the setup with a detector,
	bare chips were connected to single-sided strip detectors characterized 
	in Tables \ref{tab:cdte_detector} and \ref{tab:si_detector}, 
	where the input signals to ASICs have positive polarity.
	The interface with a computer was established 
	using the SpaceWire DIO2\cite{SpaceWire_DIO2} which contains a reconfigurable FPGA. 
	The dedicated software was written for the ASIC operation on the computer.
	\begin{figure}
		\begin{center}
			\includegraphics[width=0.75\linewidth]
			{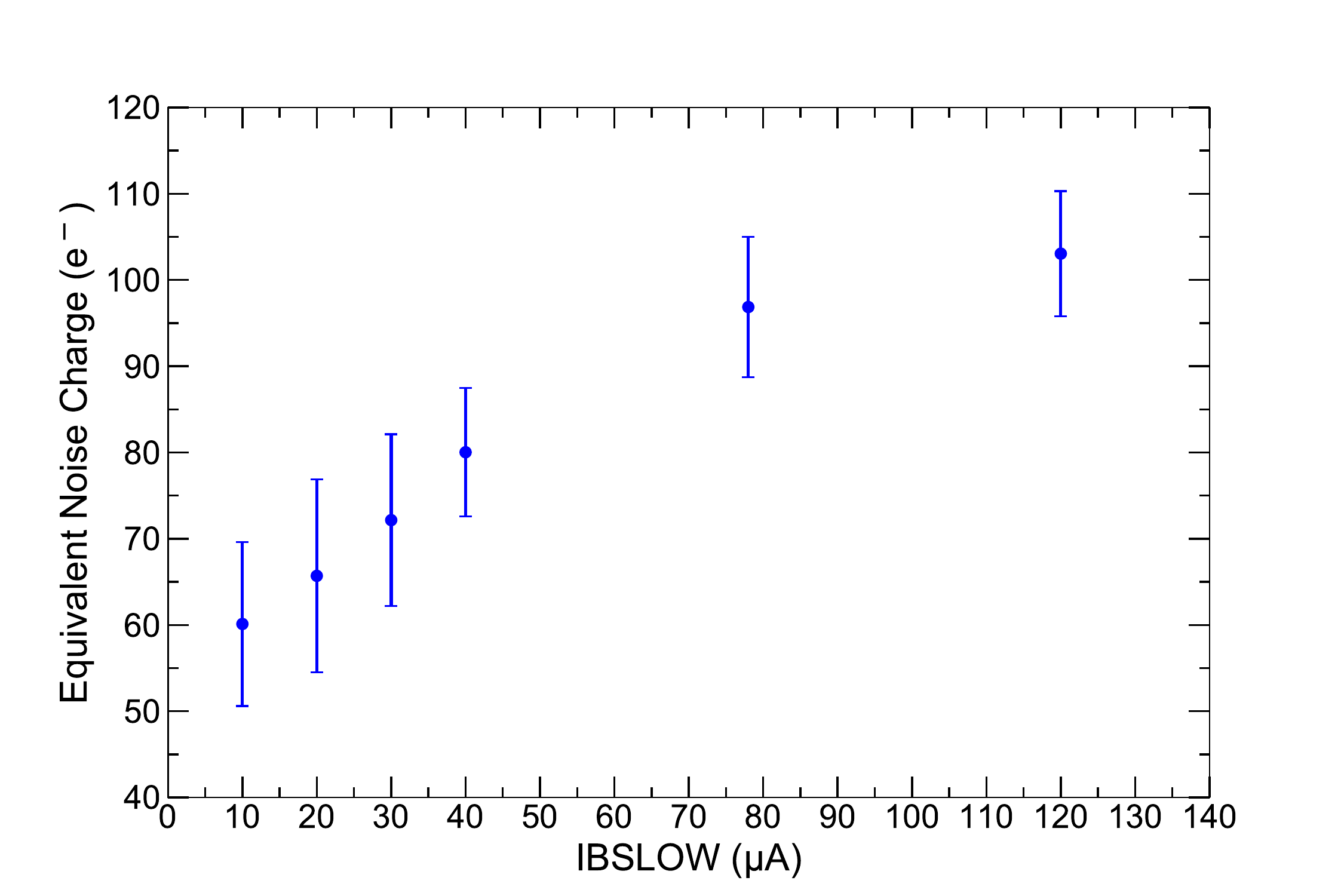}
			\caption{Noise performance at the slow shaper with respect to IBSLOW.
			The amplitude of the test pulse is $-0.93\,\si{fC}$.
			The error bars represent one-sigma among the 56 channels.} 
			\label{fig:kw04h64_enc_ibslow_exp}
		\end{center}
	\end{figure}

	\subsection{Noise Performance}
	\label{sec:noise_performance}
	Figure \ref{fig:kw04h64_enc_ibslow_exp} shows 
	the noise performance with respect to the reference current 
	for the resistance circuit at the slow shaper.
	It was measured by injecting 1000 test pulses 
	and then examining distributions of ADC values for 56 channels.
	The amplitude of the test pulses is $-0.93\,\si{fC}$ and 
	the hold timing with respect to the hit timing was kept as $3.7\,\si{\micro s}$ 
	for $\mathrm{IBSLOW} \geq 40\,\si{\micro A}$,
	while for $\mathrm{IBSLOW} = 30, 20$ and $10\,\si{\micro A}$
	it is set at $4.0, 4.5$ and $5.2\,\si{\micro s}$ 
	because of the slew-rate limited mode in the slow shaper, respectively
	(see Figure \ref{fig:kw04h64_waveform_slow_exp}).
	From the figure, 
	the noise performance is improved more sharply as IBSLOW becomes smaller 
	for $\mathrm{IBSLOW} \lesssim 40\,\si{\micro A}$ 
	than for $\mathrm{IBSLOW} \gtrsim 40\,\si{\micro A}$.
	We conjecture that the noise reduction for $\mathrm{IBSLOW} \lesssim 40\,\si{\micro A}$ 
	is related to the slew-rate limited mode of the resistance circuit.

\begin{figure}
		\begin{center}
			\includegraphics[width=0.8\linewidth]
			{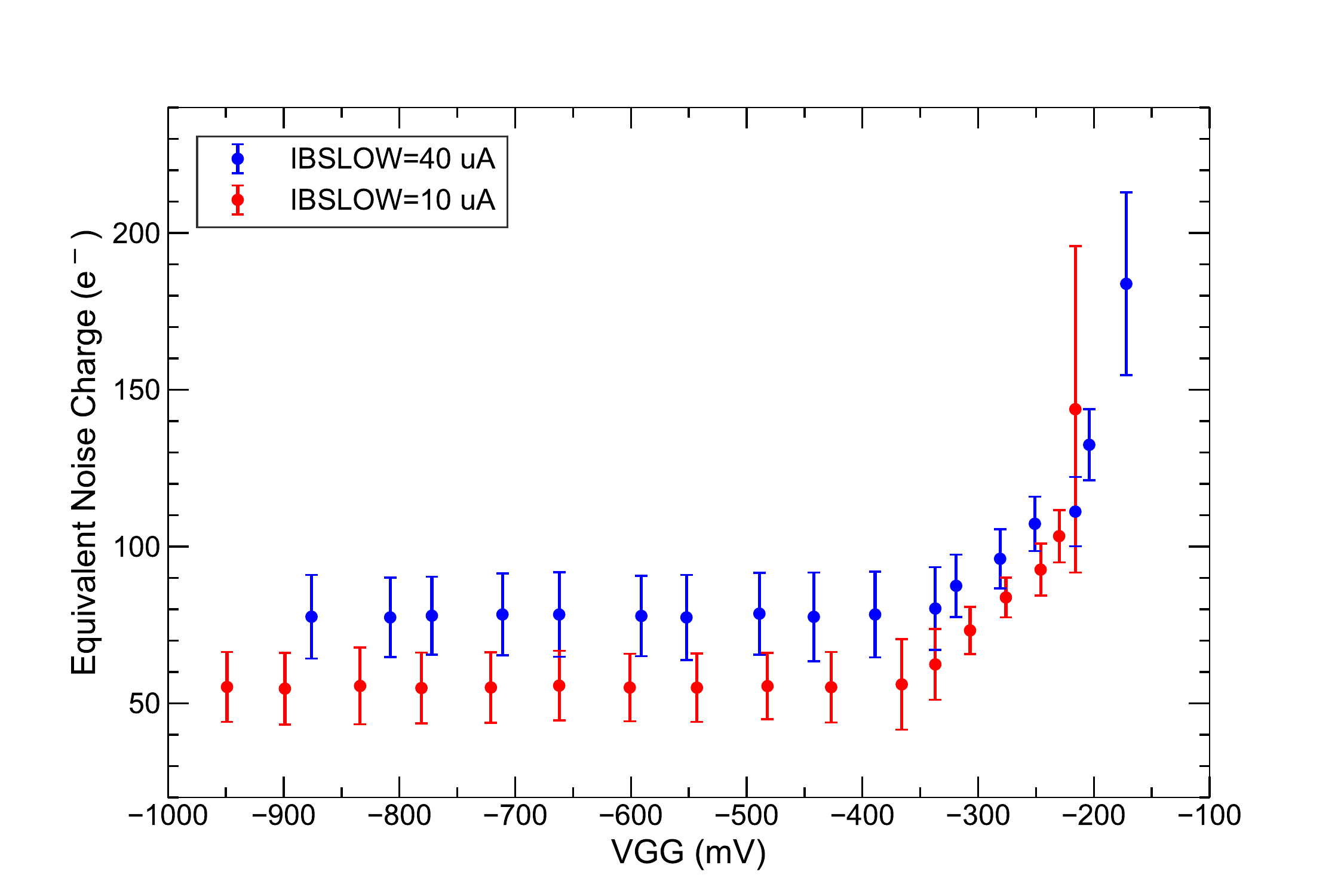}
			\caption{Noise performance at the slow shaper 
			with respect to VGG.
			The amplitude of the test pulse is $-0.93\,\si{fC}$.
			The error bars represent one-sigma among the 56 channels.}
			\label{fig:kw04h64_enc_vgg_exp}
		\end{center}
	\end{figure}

	Figure \ref{fig:kw04h64_enc_vgg_exp} shows 
	the ENC versus the gate voltage of the NMOS 
	used as the feedback resistor at the CSA 
	when $\mathrm{IBSLOW}=10$ and $40\,\si{\micro A}$.
	The feedback resistance increases along with an increase in the absolute value of the VGG, 
	which leads to better noise performance, but at the expense of longer decay constants.
	In Figure \ref{fig:kw04h64_enc_vgg_exp},
	the ENC settles down 
	when the noise from the NMOS becomes negligible 
	compared to that from other components.
	The best noise performance with the smallest 
	$\mathrm{ENC} = 54.9\,\mathrm{e}^{-} \pm 11.3\,\mathrm{e}^{-}$ (rms)
	is achieved at the lowest value of $\mathrm{IBSLOW}=10\,\si{\micro A}$.

	\subsection{Dynamic Range Performance}
	\label{sec:dynamic_range_performance}
\begin{figure}
		\begin{center}
			\includegraphics[width=0.75\linewidth]
			{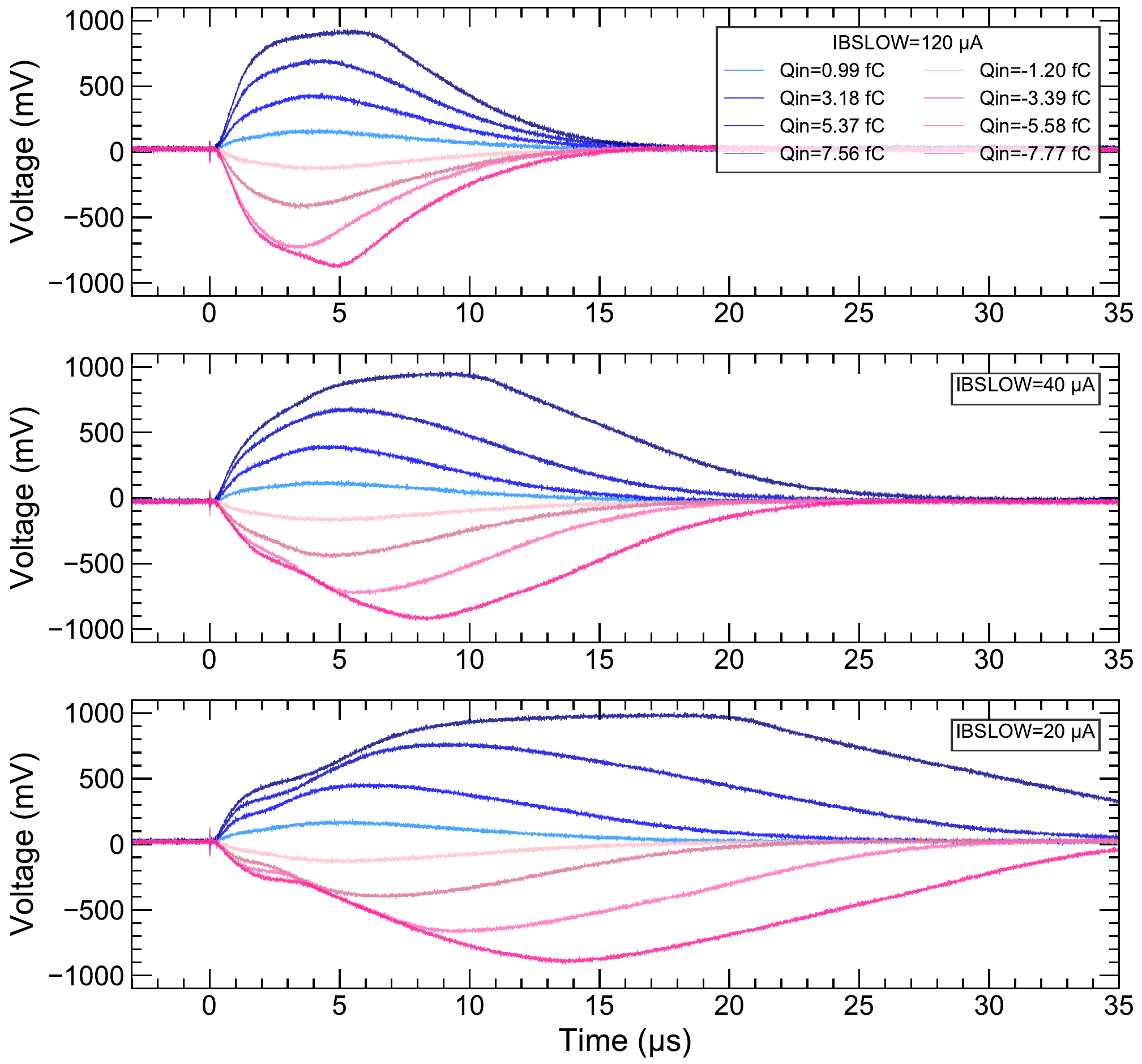}
			\caption{Waveforms at the output of the slow shaper 
			for various input charge under various IBSLOW.} 
			\label{fig:kw04h64_waveform_slow_exp}
		\end{center}
	\end{figure}

	Figure \ref{fig:kw04h64_waveform_slow_exp} shows 
	the waveforms at the slow shaper 
	for various input charge under various IBSLOW.
	The waveforms extend for smaller IBSLOW.
	This peak distortion is due to the behavior of the resistance circuit at the slow shaper.
	For smaller IBSLOW, the resistance circuit is apt to work in the slew-rate limited mode 
	and can provide small current, 
	which limits the speed of the voltage at the output of the slow shaper.
	It is also observed that negative pulses make the waveform more distorted than that of positive pulses.
	This asymmetry is also caused by the resistance circuit
	where the input NMOS shows a different behavior 
	as a response to positive and negative inputs. 

\begin{figure}
		\begin{center}
			\includegraphics[width=0.75\linewidth]
			{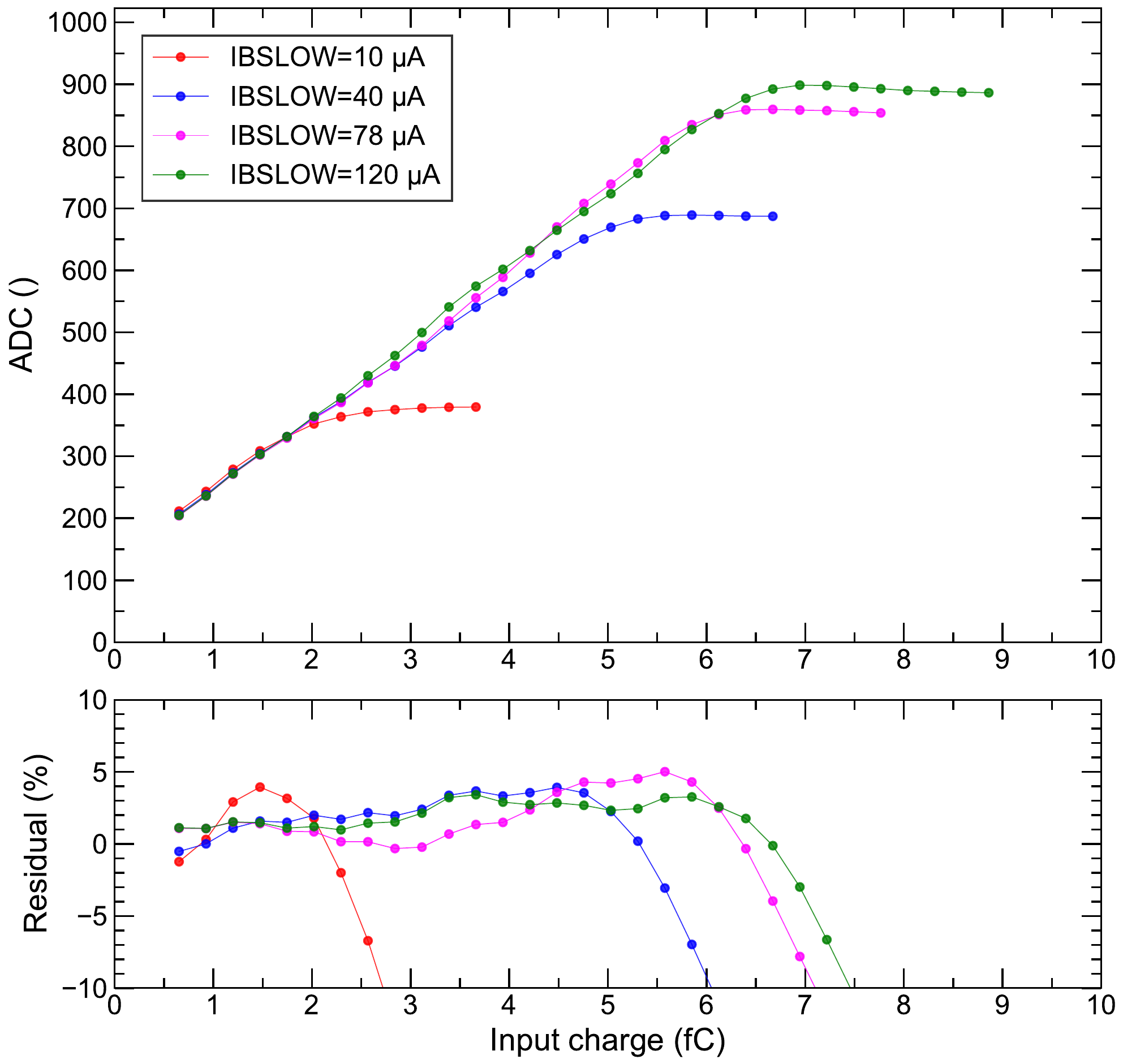}
			\caption{The upper plot shows ADC values with the input from the slow shaper 
			with respect to the input charge of a channel for different IBSLOW values.
			The data are fitted over the dynamic range for each IBSLOW value. 
			Residuals are defined as the difference of the measurement 
			divided by the saturated ADC value, and are shown in the lower plot.
			The signals with negative polarity are used.} 
			\label{fig:kw04h64_dynamic_exp_2}
		\end{center}
	\end{figure}

	Figure \ref{fig:kw04h64_dynamic_exp_2} shows 
	the relation between the input charge and the ADC value 
	representing the amplitude of the sample-and-hold voltage from the slow shaper
	for various IBSLOW. 
	The results from a typical channel are plotted.
	The hold timing with respect to the hit timing was 
	$3.7\,\si{\micro s}$ for $\mathrm{IBSLOW} \geq 40\,\si{\micro A}$ 
	and $5.2\,\si{\micro s}$ for $\mathrm{IBSLOW} =10\,\si{\micro A}$.
	The gain dispersion was $3$-$8\,\%$ in one-sigma among the 56 channels.
	The dynamic range was found to be $\sim 6.5\,\si{fC}$ at $\mathrm{IBSLOW}=120\,\si{\micro A}$ and 
	became smaller following the decrease in IBSLOW, 
	reaching $\sim 2.0\,\si{fC}$ at $\mathrm{IBSLOW}=10\,\si{\micro A}$.
	This trend can be well explained 
	by the behavior of the resistance circuit at the slow shaper 
	and the sample-and-hold circuit.
	Since in the sample-and-hold circuit the voltage is sampled 
	at a given time after the hit signal is issued, 
	it is impossible to correctly capture the peak if it moves with the signal charge.

	\subsection{Spectroscopic Performance}
	\label{sec:spectroscopic_performance}
\begin{figure*}
		\begin{center}
			\includegraphics[width=1.0\linewidth]
			{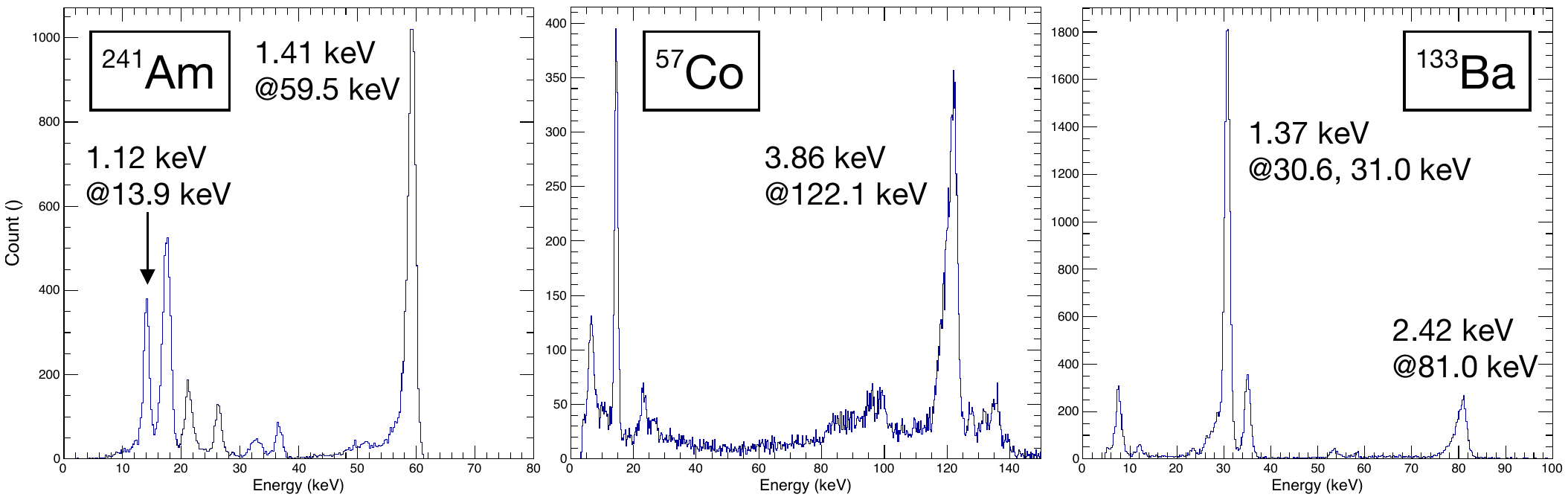}
			\caption{Energy spectra acquired with various sources 
			from one channel where
			only single hit events were extractred.} 
			\label{fig:kw04h64_cdte_spectrum_exp}
		\end{center}
	\end{figure*}

	The spectra in Figure \ref{fig:kw04h64_cdte_spectrum_exp} 
	were acquired using the CdTe detector 
	with $^{241} \mathrm{Am}$, $^{57} \mathrm{Co}$ and $^{133} \mathrm{Ba}$ sources, 
	where the detector was biased at $1000\,\si{V}$ and cooled at $-20\,\si{\degreeCelsius}$.
	The slow shaper was selected as the input of the ADC, 
	and the reference current was set at $\mathrm{IBSLOW}=40\,\si{\micro A}$.
	We confirmed the detection of photons whose energy ranges from $6.4\,\si{keV}$ to $122.1\,\si{keV}$.
	The energy resolution at $13.9\,\si{keV}$ (FWHM) was found to be $1.12\,\si{keV}$.
	We also found that energy resolution becomes worse as the peak energy increases, 
	which we will investigate further. 

\begin{figure}
		\begin{center}
			\includegraphics[width=0.80\linewidth]
			{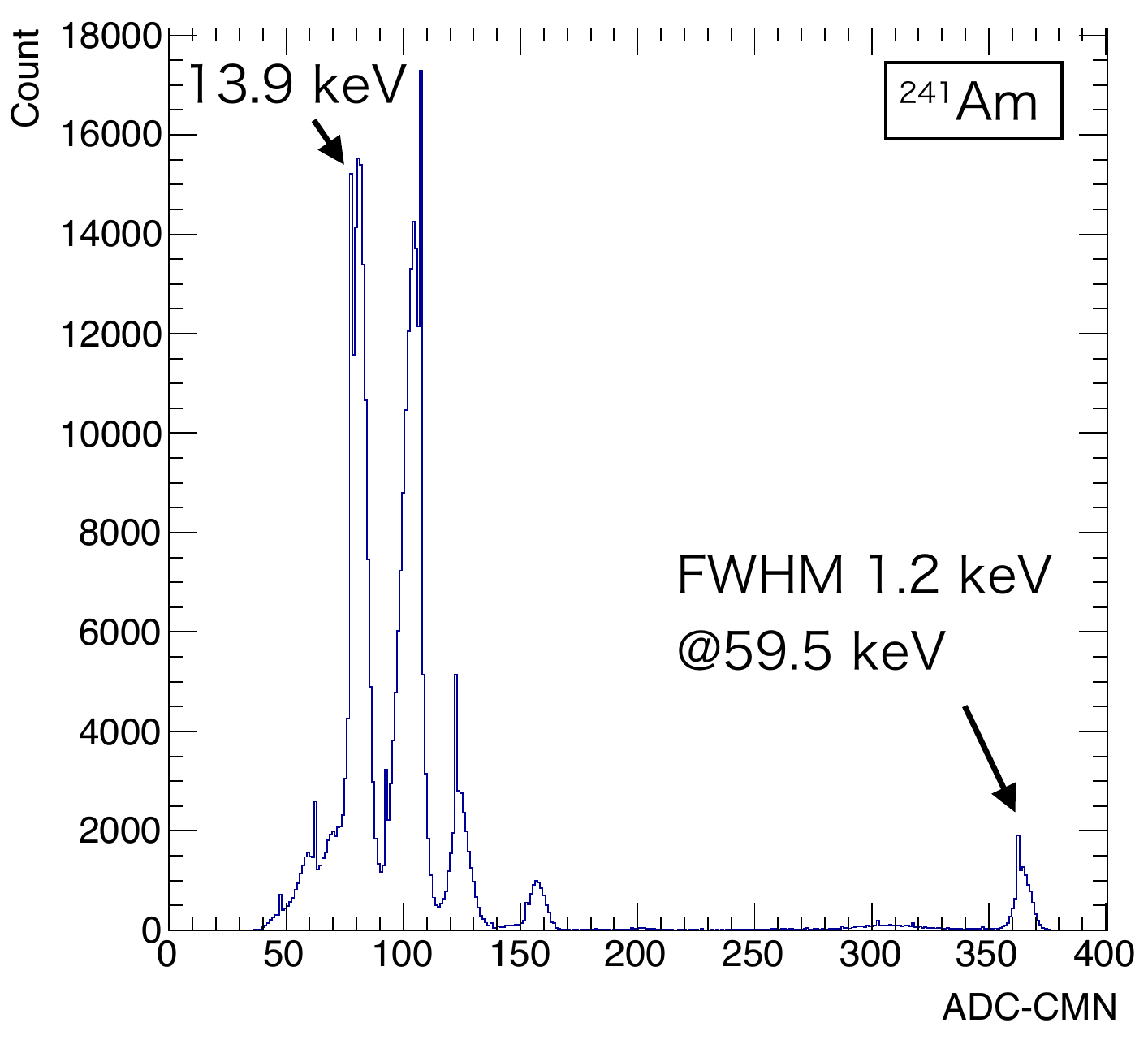}
			\caption{Energy spectrum acquired using an $^{241} \mathrm{Am}$ source 
			from one channel where
			only single hit events were extractred.
			The horizontal axis represents the value of ADC subtracted by common mode noise.} 
			\label{fig:kw04h64_si_241am_spectrum_exp}
		\end{center}
	\end{figure}

	Figure \ref{fig:kw04h64_si_241am_spectrum_exp} shows 
	the spectrum acquired using an $^{241} \mathrm{Am}$ source, 
	where the Si detector was biased at $80\,\si{V}$ and cooled at $-10\,\si{\degreeCelsius}$.
	The slow shaper was selected as the input of the ADC, 
	and the reference current was set at $\mathrm{IBSLOW}=40\,\si{\micro A}$.
	The energy resolution was found to be $1.2\,\si{keV}$ at $59.5\,\si{keV}$.

\section{New low-noise readout architecture}
\label{sec:new_low_noise_readout_architecture}
\begin{figure}
	\begin{center}
		\includegraphics[width=0.95\linewidth]
		{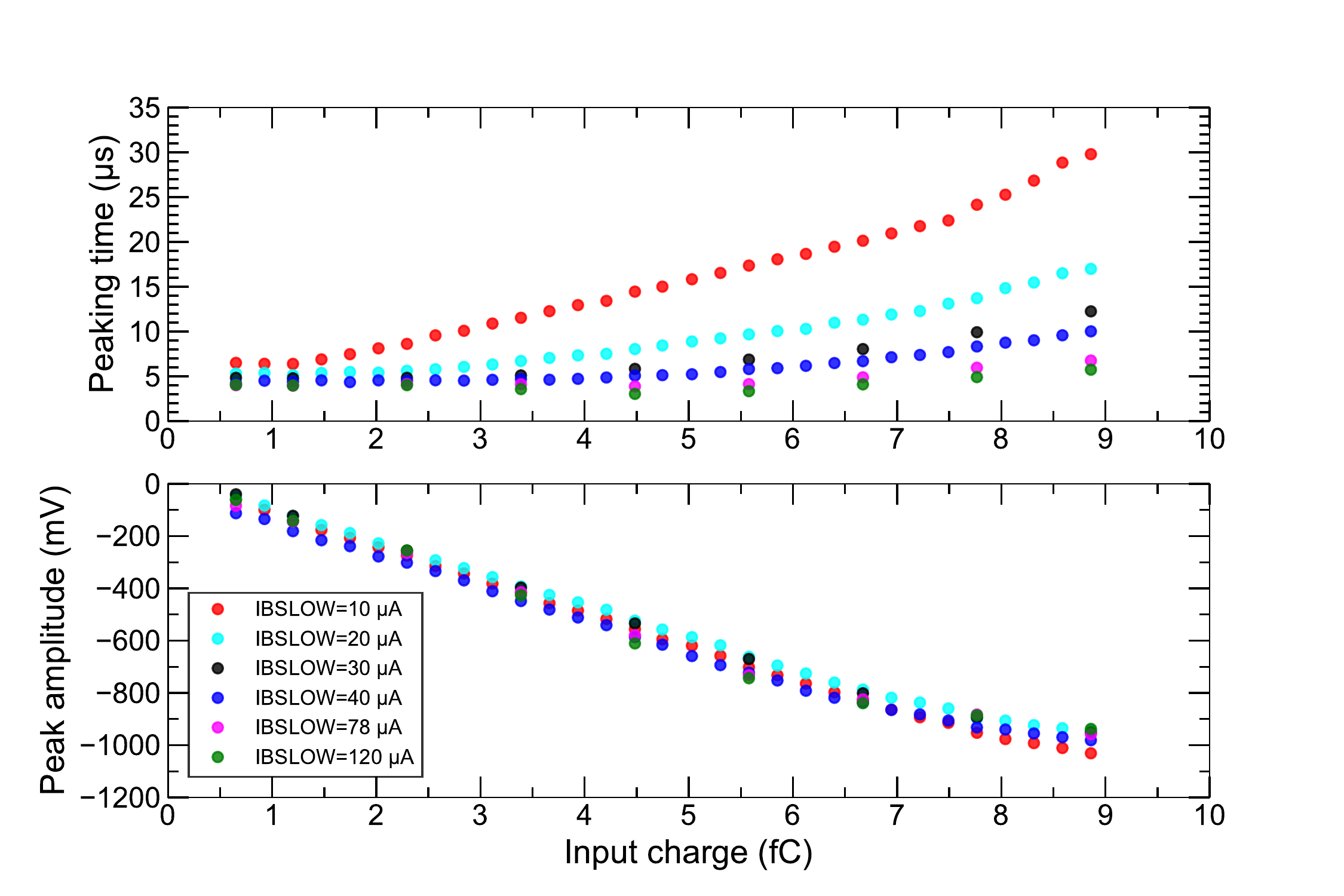}
		\caption{Peaking time and amplitude 
		with respect to the test pulse negative input charge for different IBSLOW values.
		The peaking time is defined as the time when the waveform reaches its minimum value.} 
		\label{fig:kw04h64_peak_time_voltage_exp_neg}
	\end{center}
\end{figure}

According to the results reported 
in Section \ref{sec:noise_performance} and \ref{sec:dynamic_range_performance},
the slew-rate limited mode offers better noise performance 
with some penalty in the dynamic range 
as long as the sample-and-hold circuit is employed.
Figure \ref{fig:kw04h64_peak_time_voltage_exp_neg} shows 
the peaking time and the peak amplitude with respect to the negative input charge.
The peaking time gets longer as the input charge increases 
when the slow shaper works in the slew-rate limited mode, 
while the peak voltage maintains the linearity regardless of the mode
for the input charge up to and above $8\,\si{fC}$.
This implies that 
the combination of the slew-rate limited mode and the peak detector circuit
should satisfy both the noise performance and the wide dynamic range.

On the other hand, it should be noted that 
there exist some reservations to employ the peak detector circuit with the slew-rate limited mode.
This architecture makes it hard to detect the common mode noise.
Care must be taken so that the AD conversion starts 
after the voltage reaches its peak for the largest signal to be targeted.
In addition, the peaking time reflects noise performance.
These matters must be taken into account in employing this architecture.

\section{Summary}
The KW04H64 ASIC has been designed for 
the readout of CdTe DSD 
allowing for high spectroscopic imaging.
Evaluating its performance experimentally, 
the low noise performance, ENC of $54.9\,\mathrm{e}^{-} \pm 11.3\,\mathrm{e}^{-}$ (rms), 
has been demonstrated without any detector. 
From the evaluation of the spectroscopic performance using the CdTe single-sided detector, 
the ASIC demonstrated the high energy resolution of $1.12\,\si{keV}$ for the energy at $13.9\,\si{keV}$ and 
the capability of detecting photons within the energy
from $6.4\,\si{keV}$ to $122.1\,\si{keV}$.
Investigating the circuit behavior 
under the slew-rate limited mode of the slow shaper,
we propose a new readout architecture incorporating a peak detector circuit 
to achieve a low noise readout without sacrificing the dynamic range.
The confirmation of the readout architecture 
and the performance evaluation of the CdTe DSD will follow.

\section{Acknowledgement}
This work was supported by JSPS KAKENHI grant number 18H05463.
The authors thank P. Caradonna for his critical reading of the manuscript.

\bibliography{mybibfile}
\end{document}